\documentclass[twocolumn,aps,prd,amsmath,superscriptaddress]{revtex4-1}
\usepackage{cancel}
\usepackage{graphicx}
\usepackage{color}
\usepackage{setspace}
\usepackage{fancyhdr}
\usepackage{amssymb}
\usepackage{amsmath}
\usepackage{float}
\usepackage{todonotes}
\usepackage{booktabs}
\usepackage[utf8]{inputenc}

\newcommand{\bea}{\begin{eqnarray}}
\newcommand{\eea}{\end{eqnarray}}

\begin{document}

\title{Relativistic parameterizations of neutron matter and implications for
  neutron stars}

\author{Nadine Hornick}
\affiliation{Institut f\"ur Theoretische Physik, Goethe Universit\"at,
Max-von-Laue-Stra\ss{}e 1, D-60438 Frankfurt, Germany} 
  
 \author{Laura Tolos}
\email{tolos@th.physik.uni-frankfurt.de}
\affiliation{Institut f\"ur Theoretische Physik, Goethe Universit\"at, Max-von-Laue-Stra\ss{}e 1, D-60438 Frankfurt, Germany} 
\affiliation{FIAS, Goethe Universit\"at, Ruth Moufang Str 1, D-60438 Frankfurt, Germany} 
\affiliation{Institute of Space Sciences (ICE, CSIC), Campus UAB, Carrer de Can Magrans, 08193 Barcelona, Spain}
\affiliation{Institut d'Estudis Espacials de Catalunya (IEEC), 08034 Barcelona, Spain}

\author{Andreas Zacchi}
\email{zacchi@astro.uni-frankfurt.de}
\affiliation{Institut f\"ur Theoretische Physik, Goethe Universit\"at, Max-von-Laue-Stra\ss{}e 1, D-60438 Frankfurt, Germany} 

\author{Jan-Erik Christian}
\email{christian@astro.uni-frankfurt.de}
\affiliation{Institut f\"ur Theoretische Physik, Goethe Universit\"at,
Max-von-Laue-Stra\ss{}e 1, D-60438 Frankfurt, Germany} 

\author{J\"urgen Schaffner-Bielich}
\email{schaffner@astro.uni-frankfurt.de}
\affiliation{Institut f\"ur Theoretische Physik, Goethe Universit\"at, Max-von-Laue-Stra\ss{}e 1, D-60438 Frankfurt, Germany} 

\date{\today}

\begin{abstract}
  We construct parameter sets of the relativistic mean-field model fitted to
  the recent constraints on the asymmetry energy $J$ and the slope parameter
  $L$ for pure neutron matter. We find cases of unphysical behaviour, i.e.\
  the appearance of negative pressures, for stiff parameter sets with low
  values of the effective mass $m^*/m$. In some cases the equation of state
  of pure neutron matter turns out to be outside the allowed band given by
  chiral effective field theory. The mass-radius relations of neutron stars
  for all acceptable parameter sets shows a maximum mass in excess of
  $2M_\odot$ being compatible with pulsar mass measurements. Given the constraints on the model in the low-density regime coming from chiral effective theory, we find that the radius of a $1.4M_\odot$ neutron star is nearly independent on the value of $L$. This is  in contrast to some previous
  claims for a strong connection of the slope parameter with the radius of a
  neutron star. In fact, the mass-radius relation turns out to depend only on
  the isoscalar parameters of symmetric matter. The constraints of
  GW170817 on the tidal deformability and on the radius are also discussed.
\end{abstract}

\maketitle

%%%%%%%%%%%%%%%%%%%%%%%%%%%%%%%%%%%%%%%%%%%%%%%%%%%%%%%%%%%%%%%%%%%%%%%%%%%

\section{Introduction} 

Neutron stars have received a lot of attention over the years
\cite{Lattimer:2004pg,Lattimer:2006xb,Oertel:2016bki}, especially since the
detection of gravitational waves from the neutron star binary merger GW170817
\cite{TheLIGOScientific:2017qsa}. The two most important properties of
neutron stars, maximum mass and radius, are still a matter of intense analysis since they are
linked to the physics of their interior, which is nowadays an open question.

It is now established that neutron stars, usually observed as pulsars, can have masses of up to $2M_{\odot}$ \cite{Demorest:2010bx,Antoniadis:2013pzd,Fonseca:2016tux}. The precise determination of neutron star radii is still an ongoing process. Model dependent
constraints on the radius have been derived by fits to low mass quiescent X-ray
binary data and thermonuclear bursts, sometimes with conflicting results
\cite{Verbiest:2008gy,Ozel:2010fw,Suleimanov:2010th,Lattimer:2012xj,Steiner:2012xt,
  Bogdanov:2012md,Guver:2013xa,Guillot:2013wu,Lattimer:2013hma,Poutanen:2014xqa,
  Heinke:2014xaa,Guillot:2014lla,Ozel:2015fia,Ozel:2015gia,Ozel:2016oaf,
  Lattimer:2015nhk,Steiner:2017vmg}. Recent analysis of tidal deformabilities
from the neutron star merger seen in gravitational waves, GW170817, however,
were able to set limits on the radius of a $M=1.4M_\odot$ neutron star in the
range of $R=12$--$13.5$~km based on statistical approaches
\cite{TheLIGOScientific:2017qsa,Annala:2017llu,Most:2018hfd,De:2018uhw,Abbott:2018exr}.
Future high-precision X-ray space missions, such as the on-going NICER (Neutron
star Interior Composition ExploreR) \cite{2014SPIE.9144E..20A} and the future
eXTP (enhanced X-ray Timing and Polarimetry Mission) \cite{Zhang:2016ach}, will
improve the situation by simultaneous measurements of masses and radii with
higher accuracy \cite{Watts:2016uzu}. Limits on neutron star radii are also
expected to be refined by future detections of gravitational-wave signals from
neutron star mergers.
  
The mass and radius of neutron stars strongly depend on the properties of matter in their
interior, that is described by means of the equation of state (EoS). Indeed, the determination of the EoS is a field of extensive and active research.  Among
the different approaches to obtain the EoS, the relativistic mean field (RMF)
models
\cite{Serot:1984ey,Mueller:1996pm,Serot:1997xg,Glendenning:2000,Fattoyev:2010rx,Chen:2014sca}
have been widely used for describing the interior of neutron stars based on fits
to nuclear ground state properties and/or on fitting the parameters of the model
directly to properties of nuclei, such as masses, charge radii and surface
thickness. Yet it is far from a trivial task to generate an EoS that respects
the properties of nuclear matter and nuclei as well as describes pure neutron
star matter. Recall that it is an extrapolation of $\sim$18 orders of magnitude
from the radius of a nucleus to the radius of a neutron star
\cite{Weber:1989hr,Horowitz:2000xj}. For densities
$\rho \leq 4\cdot 10^ {11}~\rm{g/cm^ 3}$ neutron stars are expected to have a
outer crust consisting of a lattice of neutron-rich nuclei. Up to densities of
about $\rho \sim 10^ {14}~\rm{g/cm^ 3}$, the inner crust consists of a lattice of
nuclei immersed within a neutron liquid. At higher densities the outer core is liquid
neutron rich matter, consisting of a liquid of neutrons with a small admixture
of protons and electrons. For the inner core, probably starting at twice
saturation density, i.e.\ at $\rho \geq 5\cdot 10^ {14}~\rm{g/cm^ 3}$, one may
speculate about exotic phases of matter, such as hyperon matter (see Ref.~\cite{Chatterjee:2015pua} for a
review) or quark matter (see e.g.\
\cite{Weber:2004kj,Alford:2006vz,Weissenborn:2011qu,Buballa:2014jta,Zacchi:2015lwa}).

In this paper we present relativistic parameter sets for the EoS of the interior
of neutron stars that fulfill the $M\geq 2M_\odot$ neutron star mass constraint
from the observations of pulsars and the radius constraint from GW170817 of
$R=12$--$13.5$~km, while fulfilling the saturation properties of nuclear
matter. Moreover, we impose further constraints on the EoS for 
neutron matter coming from chiral effective field theory ($\chi$EFT)
\cite{Drischler:2016djf}. These constraints are met by simultaneously fitting
the isoscalar couplings to saturation properties, while allowing for variations
of the isovector parameters and the effective nucleon mass ($m^*/m$) so as to
reproduce the symmetry energy ($J$) and its slope ($L$) within reasonable
theoretical and experimental limits
\cite{Li:2013ola,Lattimer:2012xj,Roca-Maza:2015eza,Hagen:2015yea,Oertel:2016bki,Birkhan:2016qkr}.

We find that the values of the symmetry energy and its slope that allow for a
physical solution for the neutron matter EoS compatible with $\chi$EFT depend on the
value of the nucleon effective masses at saturation density. We also observe that
the behaviour of both the maximum mass and the radius of neutron stars is
dominated by the effective nucleon mass. The restricted range of $L$ values
coming from $\chi$EFT constraints does not allow for an appreciable variation of
the radius, while being not relevant for the determination of the maximum
mass. All parameters sets result in maximum neutron star masses in excess of the
$2M_\odot$ limit. Large values of the effective nucleon mass induce small
neutron star radii, so that effective nucleon masses of $m^*/m > 0.60$ are
needed in order to have radii compatible with the recent upper limit of the
tidal deformabilities and radii from GW170817.  

The article is organized as follows. In Section \ref{formalism} we present the
model Lagrangian and derive the corresponding equations of motion. We determine
the parameters of the model in Section \ref{parameters}, while presenting our results
for the EoS, mass-radius relation and dimensionless tidal deformability
in Section \ref{results}. Our conclusions are summarized in Section
\ref{conclusions}. The tables with the isoscalar and isovector
parameters of the model can be found in the Appendix \ref{app}.

%%%%%%%%%%%%%%%%%%%%%%%%%%%%%%%%%%%%%%%%%%%%%%%%%%%%%%%%%%%%%%%%%%%%%%%%%%%
\section{Theoretical framework}
\label{formalism}

The properties of the nuclear EoS and of finite nuclei can be described within
the RMF model, using a contemporary formulation of the Lagrangian density of the
theory
\cite{Serot:1984ey,Mueller:1996pm,Serot:1997xg,Glendenning:2000,Fattoyev:2010rx,Chen:2014sca}.
In this framework, the interactions among nucleons (N), three mesons ($\sigma$,
$\omega$ and $\rho$) and the photon can be depicted by the effective
Lagrangian for the interaction
\begin{eqnarray} 
  \nonumber
  \mathcal{L_{\rm int}} 
  &=& \sum_{N}
      \bar \Psi_{N} \left[g_{\sigma} \sigma -
      g_{\omega}\gamma^{\mu}\omega_{\mu}-\frac{g_{\rho}}{2} \gamma^{\mu}
      \vec{\tau} \vec{\rho}_{\mu} -q_N \gamma^{\mu}A_{\mu}\right]\Psi_N \\
  \nonumber  
  &-& \frac{1}{3}b\, m (g_{\sigma}\sigma)^{3}-\frac{1}{4}c(g_{\sigma}\sigma)^{4}\\ 
  &+&\Lambda_{\omega}(g_{\rho}^{2}\vec{\rho}_{\mu}\vec{\rho\,}^{\mu})(g_{\omega}^{2}\omega_{\mu}\omega^{\mu})
      +\frac{\zeta}{4!}(g_{\omega}^{2}\omega_{\mu}
      \omega^{\mu})^{2} ,
      \label{Lagrangian} 
\end{eqnarray}
where $\Psi_N$ indicates the Dirac field for the nucleons ($n$=neutron and $p$=proton), $m$ is the average nucleon mass, while $\sigma$, $\omega_{\mu}$
and $\vec{\rho}_{\mu}$ are the mesonic fields and $A_{\mu}$ the photon
field. Also, $q_N$ is the charge of the nucleon and $\vec{\tau}$ indicates the
Pauli matrices.

The $g_{\sigma}$ and $g_{\omega}$ couplings of the isoscalar $\sigma$ and
$\omega$ mesons to the nucleon determine the energy per particle and density
of the ground state of nuclear matter.  The $g_{\rho}$ coupling of the
isovector $\rho$ meson to the nucleon is important for the nuclear symmetry
energy. Moreover, the $\sigma$-meson self-interactions, with the $b$ and $c$
couplings, allow for a quantitatively successful description of nuclear matter
and finite nuclei, as they soften the EoS at moderate densities giving rise to
a realistic incompressibility of nuclear matter
\cite{Boguta:1977xi,Boguta:1981px,Youngblood:1999zza}. The mixed interaction
among $\omega$ and $\rho$, $\Lambda_\omega$, models the density dependence of
the nuclear symmetry energy
\cite{Horowitz:2000xj,Horowitz:2001ya,Fattoyev:2010rx}, influencing the
neutron radius of heavy nuclei and, presumably, neutron star radii. Finally, the quartic
self-coupling $\zeta$ of $\omega$ softens the EoS at high densities
\cite{Mueller:1996pm,Horowitz:2000xj,ToddRutel:2005fa}, affecting the maximum
mass of neutron stars.

Given the Lagrangian density of Eq.~(\ref{Lagrangian}), one can derive the
equation of motion for each particle. Nucleons satisfy
the Dirac equation \begin{eqnarray}
&&(i\gamma^{\mu}\,\partial_{\mu}- q_N \gamma^{\mu}\,A_{\mu}-m^{*} \nonumber \\
&&-g_{\omega}\, \gamma^{0} \, \omega_{0} -\frac{g_{\rho}}{2}\, \tau_{3N}\,
\gamma^{0} \,{\rho_3}_{0}) \Psi_N=0 , \nonumber 
\end{eqnarray} with $\tau_{3p}=+1$ for the proton and the effective mass of the nucleon
being
\begin{equation} 
m^{*}=m-g_{\sigma} \sigma \label{effmass} .  
\end{equation}
While the photon obeys the Poisson equation with the proton density being the
source term, the mesonic equations of motion follow from the respective
Euler-Lagrange equations. In the mean-field approach, the mesons are replaced
by the mean-field expectation values, that is,
$\bar \sigma= \langle \sigma \rangle$, $\bar\omega=\langle\omega^0\rangle$,
$\bar\rho=\langle\rho_3^0 \rangle$. Thus, the mesonic equations of motion read
\begin{eqnarray} \label{mesonic_eoms}
  && m_\sigma^2 \, \bar \sigma + m \, b \,  g_{\sigma}^3 \bar \sigma^2 + c \,
     g_{\sigma}^4 \bar \sigma^3 =  g_{\sigma} \, n^s , \nonumber \\  
  && m_\omega^2 \, \bar \omega + \frac{\zeta}{3!}  g_{\omega}^4 \bar \omega^3
     + 2 \Lambda_{\omega} g_{\rho}^2  g_{\omega}^2  \bar \rho^2 \bar \omega =
     g_{\omega} \, n , \nonumber \\  
  && m_\rho^2 \,  \bar \rho + 2 \Lambda_{\omega} g_{\rho}^2  g_{\omega}^2
     \bar \omega^2 \bar \rho=   \frac{g_{\rho}}{2} \, n_3 \ .  
\end{eqnarray}
The quantities $n^s=n^s_p+n^s_n$ and $n=n_p+n_n$ are the scalar and vector
nuclear densities, respectively, whereas $n_3=n_p-n_n$, all of them in terms of
the proton and neutron densities, generically defined as
\begin{eqnarray}
n^s_{N}&=&\frac{m^{*}}{2 \pi^{2}} \left[E_{F_N}k_{F_N}-{m}^{*2} \ln
           \frac{k_{F_N}+E_{F_N}}{m^{*}} \right] \ , \nonumber \\ 
n_N&=&\frac{k_{F_N}^{3}}{3\pi^{2}} ,
\end{eqnarray} 
with $E_{F_N}=\sqrt{k_{F_N}^{2}+m^{*2}}$, where $k_{F_N}$ is the nucleon Fermi
momentum.

With the above results, one can derive the energy density  from the energy
momentum tensor in the mean-field approximation:
\begin{eqnarray}
\varepsilon&=&\sum_N \frac{1}{8\pi^ {2}}\bigg[k_{F_N}E_{F_N}^{3}+k_{F_N}^{3}
E_{F_N} - {m}^{*4} \ln\frac{k_{F_N}+E_{F_N}}{{m}^{*}} \bigg] \cr 
&+&\frac{1}
{2}m^{2}_{\sigma} \bar \sigma^{2}+\frac{1}{2}m^{2}_{\omega} \bar \omega^{2}+\frac{1}{2}m^{2}_{\rho} \bar \rho^{2} \cr
&+& \frac{1}{3}\, b\, m (g_{\sigma} \bar \sigma)^ 3+ \frac{1}{4} \, c (g_{\sigma} \bar \sigma)^ 4 \cr
&+& \frac{\zeta}{8} (g_{\omega} \bar \omega)^4 + 3 \Lambda_{\omega} (g_{\rho}
    g_{\omega} \bar \rho \, \bar \omega) ^2 . 
\end{eqnarray} 
The pressure can be obtained via
\begin{equation}
P=\sum_{N}\mu_{N}n_{N}-\varepsilon ,
\label{press}
\end{equation}
where the nucleonic chemical potential is given by 
\begin{eqnarray}
\mu_{N}&=& E_{F_N}+g_{\omega} \, \bar \omega+\frac{g_{\rho}}{2} \,\tau_{3N}   \, \bar \rho .
\end{eqnarray}

%%%%%%%%%%%%%%%%%%%%%%%%%%%%%%%%%%%%%%%%%%%%%%%%%%%%%%%%%%%%%%%%%%%%%%%%%%%
\section{Calibration of the Model}
\label{parameters}

% $C_{S}=\frac{g_{\sigma}m}{m_{\sigma}}$,
% $C_{v}=\frac{g_{\omega}m}{m_{\omega}}$, 
% $B= \frac{b}{g^{3}_{\sigma}m}$ and $C=\frac{c}{g^{4}_{\sigma}}$

In order to analyze the effect of the nuclear bulk properties of infinite
nuclear matter on the mass and radius of neutron stars, one has to connect the
coupling constants of the Lagrangian for the interaction in Eq.~(\ref{Lagrangian}) with
the nuclear matter properties. We follow the procedure described in
Ref.~\cite{Chen:2014sca}, where the isoscalar and isovector coupling constants
are fixed in terms of the different bulk parameters of infinite nuclear
matter.

On the one hand, the isoscalar coupling constants ($g_{\sigma}$, $g_{\omega}$,
$b$ and $c$) are determined by fixing the saturation density $n_0$ as well as the
binding energy per nucleon $E/A$, the incompressibility
coefficient $K$ and the effective nucleon mass $m^*/m$ at saturation.  In
this work we take $n_0=0.15$ fm$^{-3}$, $E/A(n_0)=-16$ MeV, while the incompressibility
is chosen to be $K(n_0)=240$~MeV, based on Ref.~\cite{Shlomo:2006} and within
the accepted values reported in \cite{Oertel:2016bki}.  Moreover, the
effective nucleon mass will be varied between $m^*/m$=0.55 and 0.75. The values
for the isoscalar coupling constants are given in Table~\ref{label111} of the
Appendix \ref{app} for the different values of the effective nucleon mass.
The remaining coupling constant $\zeta$ is set to zero, so as to produce the
stiffest possible EoS at high densities. In our calculations, we take $m=$~939 MeV,
$m_{\omega}=$~783 MeV and $m_{\sigma}=$~550 MeV for the masses
of the nucleon, $\omega$ and $\sigma$ mesons, respectively.

On the other hand, we explicitely show how the isovector couplings ($g_{\rho}$
and $\Lambda_{\omega}$) are determined as a function of the symmetry energy
$J$ and the slope of the symmetry energy at saturation $L$, summarizing the
procedure of Ref.~\cite{Chen:2014sca}. 

An analytical expression for the density-dependent symmetry energy is 
\begin{eqnarray}
  && S(n)=S_0(n) + S_1(n)= \frac{k_{F}^{2}}{6E_{F}}+\frac{g_{\rho}^{2}n}{8
     m_{\rho}^{*2}}, \nonumber \\ 
  &&{\rm with} \ \
     \frac{m_{\rho}^{*2}}{g_{\rho}^2}=\frac{m_{\rho}^2}{g_{\rho}^2}+2
     \Lambda_{\omega}g_{\omega}^{2}\bar
     \omega^2 . 
\label{ssss}
\end{eqnarray}
Here $S_0(n)$ and $S_1(n)$ are the isoscalar and the isovector parts of the density-dependent symmetry energy.  Given
that the isoscalar sector has been fixed, $S_0(n)$ and derivatives are known,
while $S_1(n)$ is unknown as $g_{\rho}$ and $\Lambda_{\omega}$ are not
determined. We can use, however, the experimental knowledge on the symmetry
energy at saturation density $J=S(n_0)$ as well as the slope
$L=3 n_0 \left( \frac{ dS}{dn} \right)_0$ to determine $g_{\rho}$ and
$\Lambda_{\omega}$.

We start by writing $J$ as
\begin{eqnarray}
J&=&S(n_0)=J_0 +J_1 , \nonumber \\
J_0&=&S_0(n_0)= \left( \frac{k_{F}^{2}}{6E_{F}} \right)_0 , \nonumber \\
J_1&=&S_1(n_0)= \left( \frac{g_{\rho}^{2}n}{8 m_{\rho}^{*2}} \right)_0 .
\label{js}
\end{eqnarray}
For a given $m^*/m$ at $n_0$, we can compute $J_0$. Then, $J_1$ is also known as we fix $J$.

Moreover, the slope parameter $L$ reads as
\begin{equation}
L=3n_0 \left( \frac{dS}{dn} \right)_0= L_{0}+L_{1} ,
\end{equation}
where 
\begin{eqnarray}
L_{0}&=&3n_{0} \left( \frac{dS_{0}}{dn} \right )_{0}= 3n_{0} \left[ \left(
         \frac{\partial S_{0}}{ \partial n} \right) + \left( \frac{\partial
         S_{0}}{ \partial m^*} \right) \left( \frac{\partial m^*}{\partial n}
         \right) \right]_0 \ \ \ , \nonumber \\ 
L_{1}&=&3n_{0} \left( \frac{dS_{1}}{dn} \right )_{0}= 3n_{0} \left[ \left(
         \frac{\partial S_{1}}{ \partial n} \right) + \left( \frac{\partial
         S_{1}}{ \partial \bar \omega} \right) \left( \frac{\partial \bar
         \omega }{\partial n} \right) \right]_0 \ \ . 
\label{Lnull}
\end{eqnarray}
After a bit of algebra, it was shown in Ref. \cite{Chen:2014sca} that
\begin{eqnarray}
L_{0}&=&J_{0}\left \{1+\frac{m^{*2}}{E_{F}^{2}}\left [1-\frac{3n}{m^{*}}\left
         (\frac{\partial m^{*}}{\partial n} \right )\right ]\right \}_0 \ , 
\end{eqnarray}
with 
\begin{eqnarray}
\left( \frac {\partial m^{*}}{\partial n} \right)= -  \frac{m^{*}}{E_{F}}
  \left[ \frac{m_{\sigma^*}^{2}}{g_{\sigma}^{2}}+ \frac{\partial
  n^{s}}{\partial m^{*}} \right] ^{-1}  \ , 
\end{eqnarray}
in terms of 
\begin{eqnarray}
  \frac{m_{\sigma^*}^{2}}{g_{\sigma}^{2}}
  &=&\frac{m_{\sigma}^{2}}{g_{\sigma}^{2}}+
      2  \, b \,m \, g_{\sigma} \, \bar
      \sigma + 3 \, c \, g_{\sigma}^2 \,
      \bar \sigma^{2}  , \\ 
  \frac{\partial n^{s}}{\partial m^{*}} 
  &=& \frac{1}{\pi^{2}} \left [ \frac{k_{F}}{E_{F}} \left ( E_{F}^{2}+2m^{*2}
      \right)-3\, m^{*2} {\rm ln} \left ( \frac{k_{F}+E_{F}}{m^{*}} \right )
      \right ] . \nonumber 
\end{eqnarray}
The quantity $L_0$ is given exclusively in terms of isoscalar parameters, so
it is known. As for $L_1$ \cite{Chen:2014sca}
\begin{eqnarray}
L_1=3J_{1} \left [ 1-32 \left ( \frac {g_{\omega}^{2}}{m_{\omega}^{*2}} \right
  ) g_{\omega} \, \bar \omega \, \Lambda_{\omega} \, J_{1} \right ]_0 ,
\label{lambdaomega}
\end{eqnarray}
with $m_{\omega}^*=m_{\omega}$, since $\zeta=0$ \cite{Chen:2014sca}.
Therefore, knowing $L_0$ and given $L$, $L_1$ is easily calculated and, hence, $\Lambda_{\omega}$.  

As for $g_{\rho}$, we use Eq.~(\ref{lambdaomega}) for $\Lambda_{\omega}$ as
well as Eqs.~(\ref{ssss}) and (\ref{js}), so we find
\begin{equation} \label{grho}
\frac{g_{\rho}^2}{m_{\rho}^2}= \left[ \frac{n}{8 J_1} - 2 \Lambda_{\omega} g_{\omega}^2 \bar \omega^2 \right]^{-1}_0 .
\end{equation}

The values for $g_{\rho}$ and $\Lambda_{\omega}$ are given in Tables
\ref{label1} and \ref{label11} of Appendix \ref{app} for $J=30$ and $J=32$,
respectively, and for different values of $L$, such that $40\leq L \leq
60$~MeV.
The values of $J$ and $L$ are taken to be compatible with estimations coming
from the analysis of a variety of nuclear data from terrestrial experiments,
astrophysical observations, and theoretical calculations
\cite{Li:2013ola,Lattimer:2012xj,Roca-Maza:2015eza,Hagen:2015yea,Oertel:2016bki,Birkhan:2016qkr}.

%%%%%%%%%%%%%%%%%%%%%%%%%%%%%%%%%%%%%%%%%%%%%%%%%%%%%%%%%%%%%%%%%%%%%%%%%%%%%%%%$

\section{Equation of State, Mass-Radius and Tidal Deformability of Neutron
  Stars}
\label{results}

\subsection{Constraints on the Equation of State}

As mentioned in the previous section, the EoS must fulfill certain nuclear
matter properties at saturation density, as well as certain constraints of $J$
and $L$. Moreover, we further restrict the possible EoSs by imposing that
their behaviour for neutron matter at densities around saturation density must be compatible with
recent low-density constraints coming from the analysis of $\chi$EFT for densities up to 1.3 $n_0$ \cite{Drischler:2016djf}. Thus, we
examine the EoSs compatible with these constraints for neutron matter by varying the effective
nucleon mass, which is a free parameter, within certain reasonable limits,
between $m^*/m=$ 0.55 and 0.75 \cite{Furnstahl:1999ff}. Needless to say, we
only allow for physical solutions, where neutron matter EoS always increases with density so
that no unstable regions are present.

%%%%%%%%%%%%%%%%%%%%%%%%%%%%%%%%%%%%%%%%%%%%%%%%%%%%%%%%%%%%%%%%%%%%%%%%%%%
\begin{figure}
\includegraphics[width=\linewidth,height=\linewidth]{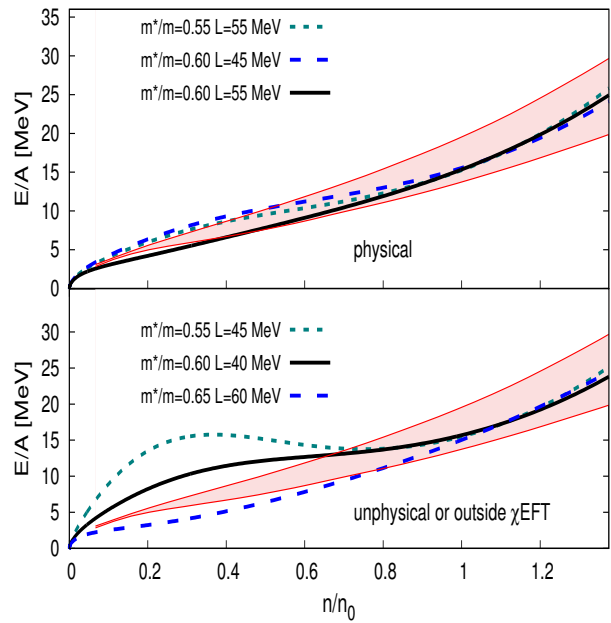}
\caption{EoS for neutron matter as a function of $n/n_0$ at $J=30$ MeV for
  different values of the nucleon effective mass $m^*/m$ and slope parameter
  $L$. The upper panel display solutions within the $\chi$EFT allowed band
  that present no instabilities, while the lower panel collects some solutions
  for the neutron matter EoS that are outside the $\chi$EFT band and/or present
  instabilities.}
\label{SlopeBand}
\end{figure}
%%%%%%%%%%%%%%%%%%%%%%%%%%%%%%%%%%%%%%%%%%%%%%%%%%%%%%%%%%%%%%%%%%%%%%%%%%%%%%%%

%%%%%%%%%%%%%%%%%%%%%%%%%%%%%%%%%%%%%%%%%%%%%%%%%%%%%%%%%%%%%%%%%%%%%%%%%%%
\begin{figure}
\includegraphics[width=\linewidth,height=\linewidth]{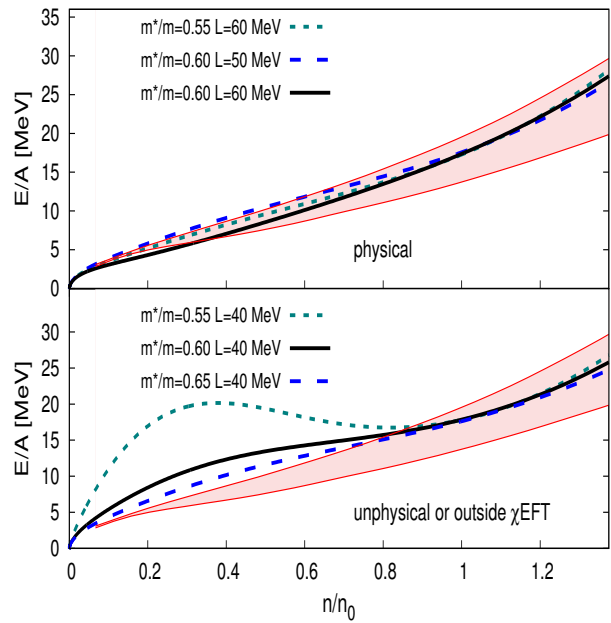}
\caption{The same as Fig.~\ref{SlopeBandErr} but for $J=32$}
\label{SlopeBandErr}
\end{figure}
%%%%%%%%%%%%%%%%%%%%%%%%%%%%%%%%%%%%%%%%%%%%%%%%%%%%%%%%%%%%%%%%%%%%%%%%%%%

With all these restrictions, we show the neutron matter EoSs for $J=30$ in
Fig.~\ref{SlopeBand} and for $J=32$ in Fig.~\ref{SlopeBandErr} by
simultaneously changing $L$ and $m^*/m$.  The upper panels in these figures
correspond to solutions of the neutron matter EoS that are compatible with the allowed band
region from $\chi$EFT (red shaded area) and present no instabilities, while
the lower panels show neutron matter EoSs that are outside the allowed $\chi$EFT and/or that are
unphysical as they present unstable regions. We observe that the physical
solutions for neutron matter EoS require larger values for $L$ as we increase $J$ for a fixed
$m^*/m$ of 0.55 MeV and 0.60 MeV.  We also find that as we take small values of
$L$, such as $L=40$ or $L=45$, a too small effective nucleon mass induces the
appearance of instabilities in the neutron matter EoS.

%%%%%%%%%%%%%%%%%%%%%%%%%%%%%%%%%%%%%%%%%%%%%%%%%%%%%%%%%%%%%%%%%%%%%%%%%%%
\begin{figure}
\includegraphics[width=\linewidth]{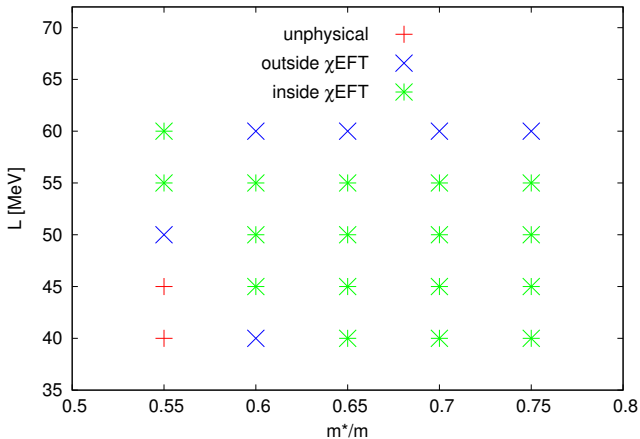}
\caption{Scatter plot of the slope parameter $L$ and of the nucleon effective mass $m^*/m$ for $J=30$ MeV, 
  where the unphysical
  solutions for neutron matter EoS (marked with a red plus sign), the solutions for neutron matter EoS
  outside the $\chi$EFT band (blue cross) and the physical solutions inside
  the $\chi$EFT region (green star) are indicated.}
\label{Slope30}
\end{figure}
%%%%%%%%%%%%%%%%%%%%%%%%%%%%%%%%%%%%%%%%%%%%%%%%%%%%%%%%%%%%%%%%%%%%%%%%%%%

%%%%%%%%%%%%%%%%%%%%%%%%%%%%%%%%%%%%%%%%%%%%%%%%%%%%%%%%%%%%%%%%%%%%%%%%%%%
\begin{figure}
\includegraphics[width=\linewidth]{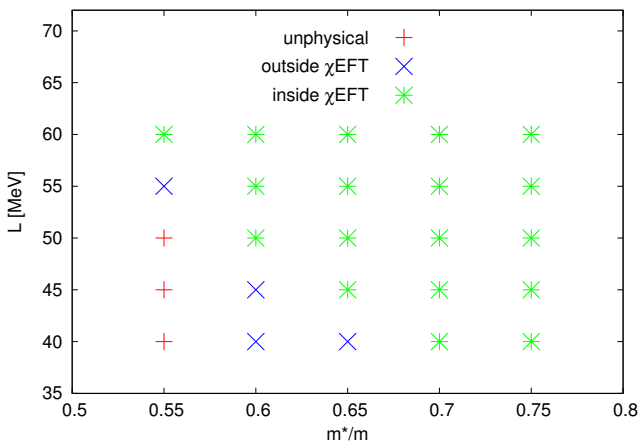}
\caption{ The same as Fig.~\ref{Slope30} but for $J=32$ MeV.}
\label{Slope32}
\end{figure}
%%%%%%%%%%%%%%%%%%%%%%%%%%%%%%%%%%%%%%%%%%%%%%%%%%%%%%%%%%%%%%%%%%%%%%%%%%%

The previous conclusions can be extracted from Figs.~\ref{Slope30} and
\ref{Slope32}, where the unphysical neutron matter EoS solutions (red plus sign), the
solutions for the neutron matter EoS outside the $\chi$EFT band (blue cross) and the physical
solutions inside the $\chi$EFT region (green star) are depicted indicating
their corresponding $L$ and $m^*/m$. These figures show a distinctive pattern
for both values of $J$: it is more difficult to find a physical solution for
the neutron matter EoS compatible with the $\chi$EFT results for small values for $m^*/m$ and
$L$ at the same time. This is due to Hugenholtz-van-Hove theorem
\cite{Boguta:1981px,Boguta:1981yn} that states that the binding energy per particle must be equal to the Fermi energy at saturation. Thus, the increase in the scalar potential (or reduction in the effective nucleon mass) will lead to bigger values for the $\omega$ field and,
hence, to a stiffer EoS. Thus, a softening of the EoS induced by a small value of
$L$ competes with the stiffening of the EoS as we reduce the effective nucleon
mass, leading to either a solution outside the band of $\chi$EFT or the
appearance of unstable solutions below saturation density. Also, we find that
it is not possible to lower the value of $J$ from 32 MeV to 30 MeV for a fixed
$m^*/m$ in order to obtain physical solutions of the neutron matter EoS that fulfill the
$\chi$EFT constraints for the whole range of $L$ studied, that is,
$40 \leq L \leq 60$, since solutions outside the $\chi$EFT emerge for a value
of $L=60$. For $J=30$ and $L=60$ it turns out that large effective nucleon
masses induce a softening of the neutron matter EoS for low-densities below the allowed
$\chi$EFT band.

\subsection{Mass and Radius of Neutron Stars}

Once the neutron matter EoS is known, we can study neutron stars by extending our analysis to $\beta$-stable neutron star matter.  Therefore,
the chemical potentials and the number densities of electrons, muons, protons and
neutrons in a neutron star core have to be related by the
conditions 
\begin{eqnarray}
\mu_n &=& \mu_p+\mu_e \ ,\nonumber \\
\mu_\mu&=&\mu_e , \nonumber \\
 n_p&=&n_e + n_\mu  .
\label{beta-eq}
\end{eqnarray} 
These relations, the Dirac equations for the nucleons (and electrons and muons), and the
field equations for the mesonic fields $\sigma$, $\omega$ and $\rho$ are to be
solved self-consistently for a given total nuclear density $n=n_p+n_n$.  Once
the chemical potential and the density of each species have been obtained at
the given $n$, one can determine the energy density and pressure (EoS) of the
neutron star matter for each density.

The mass $M$ and the corresponding radius $R$ of a non-rotating
symmetrically-spheric neutron star are obtained by solving the
Tolman-Oppenheimer-Volkoff (TOV) equations
\begin{eqnarray} \label{eq:tov_1}
  \frac{dM(r)}{dr}&=&4\pi \epsilon (r) r^2  , \nonumber \\
  \frac{dp(r)}{dr}&=&- \frac{\left[p(r)+\epsilon(r)\right] \left[M(r)+4\pi r^3
     p(r) \right]}{r(r-2M(r))}
\end{eqnarray}
in units $c=G=1$. Here $r$ is the radial coordinate and $M(r)$ is the mass
enclosed by a radius $r$.  By changing the value of the central pressure, one
obtains the mass-radius relation of neutron stars using, for the core,
the EoS described in the previous section. 

For the inner and outer crust, we use the EoS at T=0.1 MeV of the density-dependent relativistic mean-field model parameterization DD2 \cite{Typel:2009sy} in beta-equilibrium as taken from the Compose online database \cite{Compose}.
The DD2 parametrization uses density dependent couplings between the
nucleons instead of meson (self-) interactions and is fitted to properties of nuclei. 
The crust EoS considers shell effects and the lattice energy as done for the outer
crust in \cite{Baym:1971pw,Ruester:2005fm} extended to the inner
crust of a neutron star. 
The DD2 equation of state at nonzero temperature is based on the thermodynamic ansatz of \cite{Hempel:2009mc} and has been used in core-collapse supernova simulations in 
\cite{Fischer:2013eka}. Note that the crust contributes to the overall size of a 1.4$M_{\odot}$ by  $\sim 0.5$ Km.

%For the inner and outer crust, we use the supernova EoS at $T=0.1$ MeV of the density-dependent relativistic mean-field model parameterization DD2 \cite{Typel:2009sy},  adopted and used in \cite{Fischer:2013eka}, and taken from the thermodynamic ansatz of \cite{Hempel:2009mc}. The DD2 parametrization uses density dependent couplings between the nucleons instead of meson (self-) interactions and includes the formation of light clusters up to the $\alpha$ particle. Furthermore it   considers shell effects and the lattice energy as done for the outer crust in \cite{Baym:1971pw,Ruester:2005fm}, but extended to the inner crust of a neutron star. The DD2 EoS describes a smooth transition from clusterized matter as in the outer crust of a neutron star to pure nucleonic matter as present in the inner crust. Note that the crust contributes to the overall size of a 1.4$M_{\odot}$ by  $\sim 0.5$ Km.

%%%%%%%%%%%%%%%%%%%%%%%%%%%%%%%%%%%%%%%%%%%%%%%%%%%%%%%%%%%%%%%%%%%%%%%%%%%
\begin{figure}
\includegraphics[width=\linewidth]{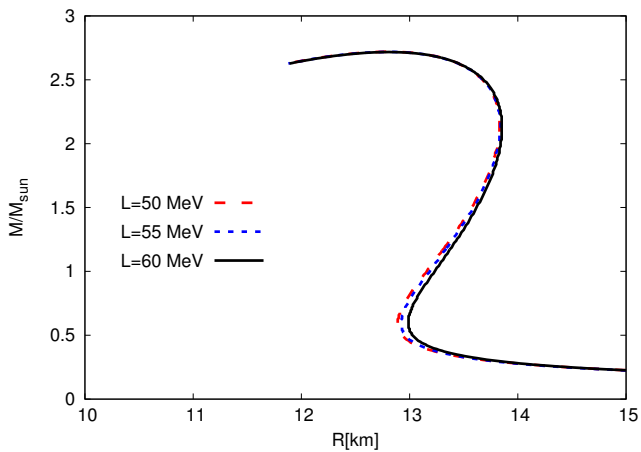}
\caption{Mass-radius relation for neutron stars for $J=32$ and a fixed value of the
  effective mass $m^*/m=0.60$, but for different values of the slope
  parameter, $50 \leq L \leq 60$. }
\label{TOVM75}
\end{figure}
%%%%%%%%%%%%%%%%%%%%%%%%%%%%%%%%%%%%%%%%%%%%%%%%%%%%%%%%%%%%%%%%%%%%%%%%%%%

Fig.~\ref{TOVM75} shows the mass-radius relation for $J=32$ and $m^*/m=0.60$ for
different values of the slope parameter $50 \leq L \leq 60$. Similar results
are found for $J=30$. We find that the 2$M_{\odot}$ limit from pulsar mass
measurements \cite{Demorest:2010bx,Antoniadis:2013pzd,Fonseca:2016tux} is fulfilled for all mass-radius curves. In fact, the maximum mass is not sensitive to
the variations of the slope parameter $L$, as already seen in
Ref.~\cite{Tolos:2016hhl,Tolos:2017lgv}. The mixed interaction between
$\omega$ and $\rho$ mesons, with the coupling $\Lambda_{\omega}$, governs the
density dependence of the nuclear symmetry energy and, hence, the slope $L$
around saturation density but turns out to be negligible at high densities, such as those found in the center of 2 $M_{\odot}$ neutron stars.
This surprising finding is due to a delicate cancellation effect. At high
densities the isoscalar vector field $\omega$ grows linearly with density, while the
equation of motion for the isovector vector field $\rho$ has a trivial solution for
a $\rho$ field growing inversely proportional to the density, i.e.\
$\rho \propto 1/n$, becoming negligible at high densities (see eqs. \ref{mesonic_eoms}).  As for the radius of a $1.4M_\odot$ neutron star, we find that it is nearly independent on the constrained value of $L$ given by $\chi$EFT ab-initio calculations. This is in contrast to some previous claims for a strong connection of the slope parameter with the radius of a neutron star which, however, consider also a larger range for $L$ \cite{Fattoyev:2012uu,Alam:2016cli,Zhu:2018ona,Lim:2018bkq}. As a consequence, all dependencies of isovector quantities on neutron star properties turn out to be small.
%Thus, we find a counterexample which shows that while the slope parameter $L$ is crucial for
%describing the neutron matter equation of state around saturation density, it
%has a negligible effect on the high-density behaviour so that the radius of
%neutron star becomes basically independent on $L$. Even more so, as the
%isovector vector field $\rho$ vanishes at high densities, all dependencies of
%isovector quantities on neutron star propeerties turn out to be nonexistent.
%We stress that this is the outcome of our particular model used and shall be
%taken as a counterexample, nothing more. 
%But at the same time it constitutes a warning for making strong claims connecting isovector observables to neutron
%star properties. 

%%%%%%%%%%%%%%%%%%%%%%%%%%%%%%%%%%%%%%%%%%%%%%%%%%%%%%%%%%%%%%%%%%%%%%%%%%%
\begin{figure}
\includegraphics[width=\linewidth]{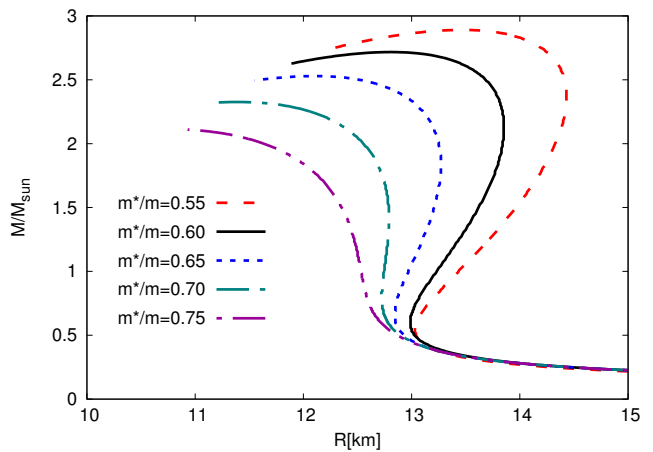}
\caption{Mass-radius relation for neutron stars for $J=32$ MeV and $L=60$ MeV for
  different values of $m^*/m$.}
\label{TOVL60}
\end{figure}
%%%%%%%%%%%%%%%%%%%%%%%%%%%%%%%%%%%%%%%%%%%%%%%%%%%%%%%%%%%%%%%%%%%%%%%%%%%

The effective nucleon mass, on the other hand, is responsible for the
variation of both the maximum mass and the radius, as seen in
Fig.~\ref{TOVL60}. Due to the Hugenholtz-van-Hove theorem, the smaller the
effective nucleon mass is, the stiffer the EoS is. Hence, small values of $m^*/m$ lead to large neutron star masses above the 2$M_{\odot}$ limit. However, recent constraints on the maximum mass coming from multi-messenger observations of GW170817 \cite{Margalit:2017dij,Rezzolla:2017aly} indicate that the maximum mass of a neutron star should be less than $\sim 2.16 M_{\odot}$. Thus, if these findings are confirmed with future observations of more GW events, this would indicate that our model for the high-dense phase is missing important contributions, such as the presence of hyperons or a phase transition that would soften the EoS and lead to the reduction of the maximum mass. With regards to the radii, small values of $m^*/m$ give rise to large radii. However, the
fact that our EoSs for neutron matter have to fulfill the $\chi$EFT constraints, that lead to
soft EoSs in the low-density regime, gives rise to radii between 12.1 and 13.7~km
for a $M=1.4M_{\odot}$ neutron star, as seen in Table~\ref{table:radius}.

%%%%%%%%%%%%%%%%%%%%%%%%%%%%%%%%%%%%%%%%%%%%%%%%%%%%%%%%%%%%%%%%%%%%%%%%%%%
\begin{figure}
\includegraphics[width=\linewidth]{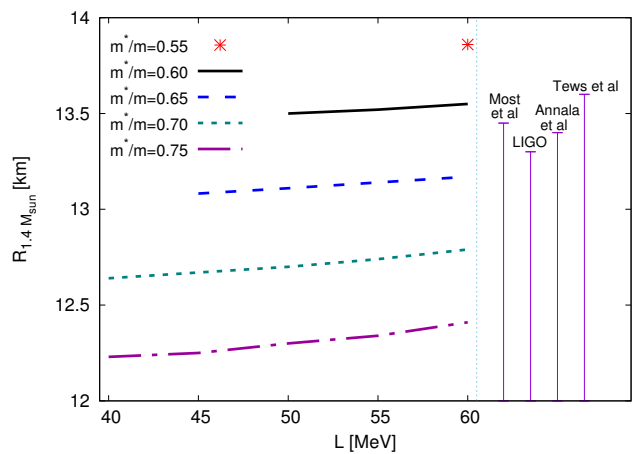}
\caption{Radius of 1.4$M_{\odot}$ neutron star as a function of the slope
  parameter $L$ for different effective nucleon masses for $J=32$ MeV.  At the
  right side of the figure, we show the upper limits on the radius of a
  $1.4M_\odot$ neutron star extracted from GW170817 by different
  groups \cite{Annala:2017llu,Most:2018hfd,Abbott:2018exr,Tews:2018iwm}.
  %Annala et al.\ \cite{Annala:2017llu} conclude that the upper limit
  %of the radius of a 1.4M$_{\odot}$ star is 13.6 km.  The work of Most et al.\
  %\cite{Most:2018hfd} finds that a purely hadronic neutron star with
  %$M=1.4M_\odot$ has a radius between
  %$12.00\leq R_{1.4M_{\odot}}({\rm km})\leq 13.45$. The recent reanalysis of LIGO
  %\cite{Abbott:2018exr} arrives at values of
  %$10.50\leq R_{1.4M_{\odot}}({\rm km})\leq 13.30$.
  %\cite{Tews:2018iwm}
  }
\label{R14Slope}
\end{figure}
%%%%%%%%%%%%%%%%%%%%%%%%%%%%%%%%%%%%%%%%%%%%%%%%%%%%%%%%%%%%%%%%%%%%%%%%%%%

In Fig.~\ref{R14Slope} we show the radius for a 1.4 $M_{\odot}$ neutron star
as a function of $L$ and $m^*/m$. We observe that changes in the effective
nucleon mass have a stronger effect on the value of the radius as compared to
variations in $L$.  As pointed out in Ref.~\cite{Alam:2016cli}, there is a
correlation of the radius of 1.4$M_{\odot}$ stars with $L$. However, the
constraints on our EoSs coming from $\chi$EFT do not allow for a wide
variation of $L$ (see Figs.~\ref{Slope30} and \ref{Slope32}). Thus, the
variation of $L$ is less important than the change of $m^*/m$ for the
determination of the radius of $1.4 M_{\odot}$ star.

\subsection{Tidal Deformability}
% {\color{blue} {\bf (all comments in blue have to be checked after calculating the tidal deformability)}}
%%%%%%%%%%%%%%%%%%%%%%%%%%%%%%%%%%%%%%%%%%%%%%%%%%%%%%%%%%%%%%%%%%%%%%%%%%%
% neue Tabelle mit Kruste und beta Gleichgewicht
\begin{table*}[ht]
\begin{center}
\begin{tabular}{|c|c|c|c|c|c|c|c|} 
\hline
$m^{*}/m$& 
$\epsilon_{c} (1.4M_{\odot})[\frac{{\rm MeV}}{{\rm fm}^3}]$ & 
$p_{c}(1.4M_{\odot})[\frac{{\rm MeV}}{{\rm fm}^3}]$ &
$R_{1.4M_{\odot}}$[km]& 
$\Lambda_{1.4M_{\odot}}$&
$\epsilon_c(2M_{\odot})[\frac{{\rm MeV}}{{\rm fm}^3}]$ &
$p_{c}(2M_{\odot})[\frac{{\rm MeV}}{{\rm fm}^3}]$ &
$M_{max}[M_{\odot}]$ \\ 
\cline{1-8}
 0.55 & 268 & 32.2  & 13.9 & 1170 & 329 & 64.2 & 2.90 \\
 0.60  & 302 & 37.0  & 13.5 & 995 & 381 & 79.0  & 2.72\\
 0.65  & 344 & 44.4  & 13.2 & 815 & 462 & 104  & 2.53\\
 0.70  & 397 & 53.4  & 12.8 & 650 & 590 & 150  & 2.32 \\
 0.75  & 466 & 65.7  & 12.4 & 501 & 856 & 261  & 2.11
\\ \hline
\end{tabular}
\caption{Central energy density, central pressure, radius and 
  dimensionless tidal deformability of a $1.4M_{\odot}$ neutron star for
  different values of $m^*/m$  and for a slope parameter of $L=60$ and a symmetry
  energy of $J=32$. We also show the central energy
  density and the central pressure  for a $2M_{\odot}$ neutron star as well as  the maximum mass for the different values of the nucleon effective mass $m^*/m$ and for $J=32$ and $L=60$.}
\label{table:radius}
\end{center}
\end{table*}
%&&&&&&&&&&&&&&&&&&&&&&&&&&&&&&&&&&&&&&&&&&&&&&&&&&&&&&&&&&&&&&&&&&&&&&&&&&&&&&&&&&&&&&&&&&&&&&&&&&&&&&&&&
%%%%%%%%%%%%%%%%%%%%%%%%%%%%%%%%%%%%%%%%%%%%%%%%%%%%%%%%%%%%%%%%%%%%%%%%%%%
% ``alte Werte`` nur mit Neutronenmaterie
% \begin{table*}[ht]
% \begin{center}
% % \renewcommand{\arraystretch}{1.75}
% \begin{tabular}{|c|c|c|c|c|c|c|c|} 
% \hline
% $m^{*}/m$& 
% $p_{c}(1.4M_{\odot})[\frac{MeV}{fm^3}]$ & 
% $\epsilon_{c} (1.4M_{\odot})[\frac{MeV}{fm^3}]$ &
% $R_{1.4M_{\odot}}$[km]& 
% $\Lambda_{1.4M_{\odot}}$&
% $p_{c}(2M_{\odot})[\frac{MeV}{fm^3}]$ & 
% $\epsilon_c(2M_{\odot})[\frac{MeV}{fm^3}]$ &
% $M_{max}[M_{\odot}]$ \\ 
% \cline{1-8}
%  0.55  & 29.57 & 257.14 & 13.71 & 869.90 & 59.04 & 316.11 & 2.91 \\
%  0.60  & 33.84 & 285.71 & 13.43 & 772.29 & 72.77 & 364.20 & 2.73\\
%  0.65  & 40.15 & 326.16 & 13.08 & 650.06 & 95.05 & 439.00 & 2.55\\
%  0.70  & 48.25 & 377.22 & 12.75 & 554.42 & 136.00 & 561.12 & 2.35 \\
%  0.75  & 59.72 & 444.65 & 12.33 & 456.70 & 235.00 & 806.05 & 2.13
% \\ \hline
% \end{tabular}
% \caption{Radius, central pressure, central energy density and weighted
% dimensionless tidal deformability of a 1.4 $M_{\odot}$ neutron star as
% function of $m^*/m$ for slope parameter of $L=60$ and symmetry energy
% $J=32$. We also show the maximum mass as well as the central pressure and
% central energy density for a 2 $M_{\odot}$ neutron star as a function of
% $m^*/m$ for the same slope paramete r $L=60$ and symmetry energy $J=32$.}
% \label{label1111}
% \end{center}
% \end{table*}
% %&&&&&&&&&&&&&&&&&&&&&&&&&&&&&&&&&&&&&&&&&&&&&&&&&&&&&&&&&&&&&&&&&&&&&&&&&&&&&&&&&&&&&&&&&&&&&&&&&&&&&&&

The recent detection of the gravitational wave event GW170817 coming from the
binary neutron star merger by the LIGO and Virgo observatories
\cite{TheLIGOScientific:2017qsa} 
has posed constraints on the EoS
through the tidal distortion during the inspiral phase. In a neutron star
binary system, the tidal deformability $\lambda$ measures the induced
quadrupole moment, $Q_{ij}$, of a star in response to the tidal field of the
companion, $\mathcal{E}_{ij}$ \cite{Hinderer:2007mb,Hinderer:2009ca}
\begin{equation}
Q_{ij}=-\lambda \mathcal{E}_{ij} .
\end{equation}
The tidal deformability $\lambda$ is related to the dimensionless $l=2$ tidal Love number $k_2$ as
\begin{equation}
 k_2=\frac{3}{2}\lambda R^{-5} , 
\end{equation}
where $R$ is the radius of the star. The tidal Love number can be calculated from
\begin{eqnarray}
k_2&=&\frac{8C^5}{5} (1-2C)^2 [2+2C(y_R-1)-y_R] \times \nonumber \\
&& \{ 2C[6-3y_R+3C(5y_R-8)]+\nonumber \\
&&4C^3[13-11y_R+C(3y_R-2)+ 2C^2(1+y_R)] + \nonumber \\
&&3 (1-2C)^2[2-y_R+2C(y_R-1)] {\rm ln}(1-2C) \}^{-1}  , \ \ \ \ \ \
\end{eqnarray}
with $C$ being the compactness parameter ($C=M/R$). The quantity
$y_R\equiv y(R)$ is obtained by solving, together with the TOV equations (\ref{eq:tov_1}),  the following differential equation for y \cite{Postnikov:2010yn}
\begin{eqnarray}
 &&ry'(r)+y(r)^2+\nonumber \\ 
 && y(r)e^{\lambda(r)}\left[1+4\pi r^2(p(r)-\epsilon(r) )\right]
 +r^2 Q(r)=0 ,
 \label{y}
\end{eqnarray}
where $e^{\lambda(r)}=\left(1-\frac{2M(r)}{r}\right)^{-1}$ is a metric function of a spherical star and 
\begin{eqnarray}
 Q(r)&=&4\pi e^{\lambda(r)}\left(5\epsilon(r)+9p(r)+\frac{\epsilon(r)+p(r)}{c_s(r)^2} \right)\\ \nonumber
 &-&6e^{\lambda(r)} r^{-2}-(\nu'(r))^2 ,
\end{eqnarray}
with $\nu'(r)$ being the derivative of the $\nu(r)$ metric function 
\begin{equation}
\nu'(r)=2e^{\lambda(r)}\frac{M(r)+4\pi p(r)r^3}{r^2} ,
\end{equation}
and $c_s(r)^2=dp/d\epsilon$ the squared speed of sound. The boundary condition of Eq.~(\ref{y}) is $y(0)$=2.

Once $k_2$ is known, the dimensionless tidal deformability $\Lambda$ can then
be determined by the relation
\begin{equation}
\Lambda=\frac{2k_2}{3C^5} .
\end{equation}

%For a binary neutron star system of masses $M_1$ and $M_2$ the weighted
%dimensionless tidal deformability is defined as
%\cite{TheLIGOScientific:2017qsa}
%\begin{eqnarray}
%\tilde \Lambda&=&\frac{8}{13} [(1+7 \eta -31 \eta^2) (\Lambda_1 + \Lambda_2) + \sqrt{1-4 \eta} \nonumber \\
% && \times (1+ 9 \eta -11 \eta^2)(\Lambda_1 -\Lambda_2) ,
%\end{eqnarray}
%where $\eta=M_1 M_2/M^2$ is the symmetric mass ratio, $M=M_1+M_2$ is the total
%mass, and $\Lambda_1$ and $\Lambda_2$ are the dimensionless tidal
%deformabilities of the binary system, for $M_1 \ge M_2$. We take
%$M_1=M_2=1.4M_{\odot}$.

In Table \ref{table:radius} we show the dimensionless tidal deformability $\Lambda$ for different values of $m^*/m$
fixing $J=32$~MeV and $L=60$~MeV as in Fig.~\ref{TOVL60}. Our value of $\Lambda$
for  $m^*/m > 0.65$ lie below the upper limit of ${\Lambda} \sim 800$ as given by
\cite{TheLIGOScientific:2017qsa}. We note that a recent reanalysis gives a
somewhat lower limit of ${\Lambda} \lesssim 500$ within 90\%
credibility \cite{De:2018uhw,Abbott:2018exr}, 
so that only the case $m^{*}/m=0.75$ would be compatible 
% allowing all our cases to be compatible
with GW170817. For completeness, we also show the radius, central energy
density and pressure of a 1.4$M_{\odot}$ neutron star, as well as the central
energy density and pressure at 2$M_{\odot}$ together with the maximum mass
reached for each value of $m^*/m$.

The constraints on the tidal deformability have also allowed for the
determination of the statistically most probable radius of a 1.4 $M_{\odot}$
neutron star. Annala et
al \cite{Annala:2017llu} has recently concluded that the maximum radius of a
1.4M$_{\odot}$ star is 13.60 km by combining the new stringent limits on tidal
deformabilities with the existence of 2$M_{\odot}$ stars using a a generic
family of EoSs that interpolate between state-of-the-art theoretical results
at low- and high- baryon density. Most et al.\ \cite{Most:2018hfd} have shown that a purely
hadronic neutron star has $12.00\leq R_{1.4M_{\odot}}({\rm km})\leq 13.45$
with a 2-$\sigma$ confidence level, with a most likely value of 12.39 km, by
imposing constraints on the maximum mass ($M<2.16 M_{\odot}$) and on the
dimensionless tidal deformability ($\Lambda < 800$ MeV) using one million
EoSs.  Also, the recent article by the LIGO and Virgo scientific collaboration
\cite{Abbott:2018exr} has claimed that
$10.50\leq R_{1.4M_{\odot}}({\rm km})\leq 13.30$ for EoSs which support
maximum masses of $M \geq 1.97 M_{\odot}$ at 90$\%$ credible level.  In Ref.~\cite{De:2018uhw} a mass range of a
binary neutron star merger with $1.1 M_{\odot} \leq M \leq 1.6 M_{\odot}$ has
been studied, leading to $8.70\leq R_{1.4M_{\odot}}({\rm km})\leq 14.10$, with
an average value of $R=11.5$~km by means of a Bayesian parameter estimation, while in \cite{Tews:2018iwm} it was found that GW170817 requires a radius of $R_{1.4M_{\odot}}< 13.6 \ {\rm km}$.
Our results for radii of a 1.4M$_{\odot}$ star are compatible with all these
analysis provided that  $m^*/m > 0.60$, as seen in Fig.~\ref{R14Slope}. Also
similar results for $R_{1.4M_{\odot}}$ are found in theoretical analysis using
different non-relativistic and relativistic models constrained by the tidal
deformabilities \cite{Kumar:2017wqp,Fattoyev:2017jql,Malik:2018zcf}.  We note
that the lower limit on the radius found in the previous works is also fulfilled
for all the cases studied in this paper.

In order to have radii below 13.5~km, we find that effective nucleon masses
$m^*/m > 0.60$ are needed. Values of the effective mass above $m^*/m=0.65$ have to be, though,
taken with caution, as these values might not
be compatible with experimental results on binding energies and charge radii
of atomic nuclei \cite{Reinhard:1989zi}.  The solution to this problem relies on the implementation of an isoscalar tensor term that will allow for effective masses $m^*/m > 0.60$ to be compatible with, for example, the large spin-orbit splitting of $^{16}{\rm O}$, which determines the magic numbers observed from nuclear binding \cite{Furnstahl:1999ff}. This term is not relevant for the present relativistic Hartree mean-field calculation \cite{Sagawa:2014cxa} of pure neutron matter and neutron star matter, but it would be of upmost relevance for the properties of finite nuclei, such as binding energies, charge radii and spin-orbit splitting, as seen in Fig. 2 of Ref.~\cite{Furnstahl:1999ff}. Work along this line is beyond the scope of the present paper and is left for a future follow-up study.

%%%%%%%%%%%%%%%%%%%%%%%%%%%%%%%%%%%%%%%%%%%%%%%%%%%%%%%%%%%%%%%%%%%%%%%%%%%%%%%%

\section{Conclusions}
\label{conclusions}

We have studied the EoS in the inner core of neutron stars within the
relativistic mean field theory with the aim of fulfilling several recent
astrophysical constraints, such as, the $2M_\odot$ neutron star mass
limit \cite{Demorest:2010bx,Antoniadis:2013pzd,Fonseca:2016tux} and the
extraction of neutron star radii  $\lesssim$ 13.5~km
from the recent analysis on tidal deformabilities of the
GW170817 neutron star merger event
\cite{Abbott:2018exr,Most:2018hfd,Annala:2017llu,De:2018uhw,Kumar:2017wqp,Fattoyev:2017jql,Malik:2018zcf}.

The phenomenological model satisfies the saturation properties of nuclear
matter together with constraints on low-density neutron matter coming from
$\chi$EFT ab-initio approaches \cite{Drischler:2016djf}. These constraints are
fulfilled by simultaneously fitting the isoscalar couplings to saturation
properties (saturation density, energy per particle and compressibility),
while allowing for variations in the isovector parameters so as to reproduce
the symmetry energy and its slope within reasonable theoretical and
experimental limits
\cite{Li:2013ola,Lattimer:2012xj,Roca-Maza:2015eza,Hagen:2015yea,Oertel:2016bki,Birkhan:2016qkr}.

We have found that the values of the symmetry energy ($30 \leq J \,\, [{\rm MeV}] \leq 32$) and
its slope ($40 \leq L \,\, [{\rm MeV}] \leq 60$) that allow for a physical solution for the neutron matter EoS
compatible with the $\chi$EFT are determined by the value of the nucleon
effective mass ($0.55 \leq m^*/m \leq 0.75$). It is indeed difficult to find a
physical solution compatible with the $\chi$EFT results once the values for
$m^*/m$ and $L$ are reduced (increased) simultaneously for a fixed value of
$J$. A softening (hardening) of the EoS induced by a small (big) value of $L$
competes with the stiffening (softening) of the EoS as we reduce (increase)
the effective nucleon mass, leading to either a solution outside the allowed
area from $\chi$EFT or the appearance of unstable solutions below saturation
density.

With regards to the mass and radius of neutron stars, we have obtained that
the effective nucleon mass turns out to be the dominant parameter controling
both, the maximum mass and the radius of a neutron star. This is due to the fact
that, on the one hand, the restricted range of $L$ values coming from $\chi$EFT constraints does
not allow for noticeable variations on the radius and, on the other hand, the isovector
parameters turn out to be not relevant
for the determination of the maximum mass, as seen in
Ref.~\cite{Tolos:2016hhl,Tolos:2017lgv}.  Large values of $m^*/m$ induce small
masses and radii, as expected from the Hugenholtz-van-Hove theorem. Thus,
effective nucleon masses of $m^*/m > 0.6$ are needed in order to reproduce
2$M_{\odot}$ observations and have radii compatible with recent astrophysical
determinations
\cite{Guillot:2013wu,Guillot:2014lla,Guver:2013xa,Heinke:2014xaa,Lattimer:2014sga,Lattimer:2013hma,Nattila:2015jra,Ozel:2015fia,Ozel:2016oaf,Lattimer:2015nhk},
that are corroborated by the analysis on tidal deformabilities of the GW170817
event
\cite{Abbott:2018exr,Most:2018hfd,Annala:2017llu,De:2018uhw,Kumar:2017wqp,Fattoyev:2017jql,Malik:2018zcf}.
In fact, our values of the dimensionless tidal deformability are
within the 90$\%$ confidence level for $m^*/m > 0.65$ at $J=32$ and
$L=60$. Note, however, that the effective nucleon mass has to be reconciled
with the binding energies and charge radius of atomic nuclei
\cite{Reinhard:1989zi}.

In the near future, apart from the expected detection of gravitational-wave
events from other neutron-star binary systems, high-precision X-ray space
missions, such as the on-going NICER \cite{2014SPIE.9144E..20A} and the eXTP
\cite{Zhang:2016ach}, will shed some more light on the properties of matter inside
neutron stars by offering simultaneous measurements of their masses and radii
\cite{Watts:2016uzu}.

%%%%%%%%%%%%%%%%%%%%%%%%%%%%%%%%%%%%%%%%%%%%%%%%%%%%%%%%%%%%%%%%%%%%%%%%%%%%%%%%

\appendix
\section{Isoscalar and isovector coupling constants}
\label{app}

In this Appendix we show the values of the coupling constants in the isoscalar
and isovector sectors used throughout this work. The isoscalar coupling
constants ($g_{\sigma}$, $g_{\omega}$, $b$ and $c$) are given in
Table~\ref{label111} for the different values of  $m^*/m$, ranging from 0.55
to 0.75.  We take $m=$~939 MeV, $m_{\omega}=$~783 MeV and $m_{\sigma}=$~550
MeV for the masses of the nucleon, $\omega$ and $\sigma$ mesons,
respectively. The coupling constant $\zeta$ is set to zero.
 
%&&&&&&&&&&&&&&&&&&&&&&&&&&&&&&&&&&&&&&&&&&&&&&&&&&&&&&&&&&&&&&&&&&&&&&&&&&&&&&&&&&&&&&&&&&&&&&&&&&&&&&&&&
%%%%%%%%%%%%%%%%%%%%%%%%%%%%%%%%%%%%%%%%%%%%%%%%%%%%%%%%%%%%%%%%%%%%%%%%%%%
\begin{table}[h]
\begin{center}
\begin{tabular}{|c|c|c|c|c|}
\hline
% \multicolumn{1}{|c|}{J=32~MeV} & \multicolumn{2}{c}{Just the physical ones...} & \multicolumn{2}{c|}{} \\
% \hline
 \hspace{.2cm} $m^{*}/m$ \hspace{.2cm} & \hspace{.1cm} $g_{\sigma}$
\hspace{.1cm} & \hspace{.2cm} $g_{\omega}$ \hspace{.2cm} & \hspace{.1cm}
b \hspace{.1cm} & c \\
\cline{1-5}
% \cline{2-5}
%  0.50 & 0.60 & 12.7990 & 0.0449062    & \\
% \cline{2-6}
 0.55 & 11.5529 & 13.5663 & -0.00198839   & -0.0028455 \\
%   Cs: 389.033 Cv: 264.349
 0.60 & 10.9935 & 12.7084 & -0.00237445  & -0.00315763\\
%  Cs: 352.269 Cv: 232.267
 0.65 & 10.4291 & 11.7742 & -0.0030843  & -0.00368166 \\
% Cs: 317.031 Cv: 199.12
 0.70 & 9.84608 & 10.7467 & -0.00431483  & -0.00434675  \\
  0.75 & 9.22731 & 9.59842  & -0.00638591  & -0.00426853 

% \cline{2-6}
% \cline{2-6}
\\ \hline
% \hline
\end{tabular}
\caption{Isoscalar coupling constants $g_{\sigma}$, $g_{\omega}$, $b$ and $c$
  for different values of $m^*/m$, ranging from 0.55 to 0.75.}
\label{label111}
\end{center}
\end{table}
%&&&&&&&&&&&&&&&&&&&&&&&&&&&&&&&&&&&&&&&&&&&&&&&&&&&&&&&&&&&&&&&&&&&&&&&&&&&&&&&&&&&&&&&&&&&&&&&&&&&&&&&&&
%&&&&&&&&&&&&&&&&&&&&&&&&&&&&&&&&&&&&&&&&&&&&&&&&&&&&&&&&&&&&&&&&&&&&&&&&&&&&&&&&&&&&&&&&&&&&&&&&&&&&&&&&&

The isovector coupling constants $g_{\rho}$ and $\Lambda_{\omega}$ are shown
in Tables \ref{label1} and \ref{label11}  for $J=30$~MeV and $J=32$~MeV, respectively, and
for different values of $L$, such that $40\leq L \,\, [{\rm MeV}] \leq 60$~MeV.  In these two
tables we also show whether the corresponding EoS is physical and inside the
allowed area coming from $\chi$EFT constraints (marked as \textcolor{green}{$\ast$}), or
unphysical (indicated as \textcolor{red}{+}) or outside the $\chi$EFT band
(depicted by \textcolor{blue}{$\times$}).

%&&&&&&&&&&&&&&&&&&&&&&&&&&&&&&&&&&&&&&&&&&&&&&&&&&&&&&&&&&&&&&&&&&&&&&&&&&&&&&&&&&&&&&&&&&&&&&&&&&&&&&&&&
%%%%%%%%%%%%%%%%%%%%%%%%%%%%%%%%%%%%%%%%%%%%%%%%%%%%%%%%%%%%%%%%%%%%%%%%%%%
\begin{table}[H]
\begin{center}
\begin{tabular}{|c|c|c|c|c|}
\hline
 \hspace{.2cm} L [MeV] \hspace{.2cm} & \hspace{.1cm} $m^{*}/m$ 
\hspace{.1cm} & \hspace{.2cm} $g_{\rho}$ \hspace{.2cm} & \hspace{.1cm}
$\Lambda_{\omega}$ \hspace{.1cm} & Solution\\
\cline{1-5}
 40 & 0.55 & 34.168 & 0.0532996 & \textcolor{red}{+}\\
 40 & 0.60 & 14.5669 & 0.0493023 & \textcolor{blue}{$\times$} \\
 40 & 0.65 & 12.4947 & 0.0465416  & \textcolor{green}{$\ast$}\\
 40 & 0.70 & 11.3062 & 0.0581025 &\textcolor{green}{$\ast$} \\
 40 & 0.75 & 10.8953 & 0.0754378 & \textcolor{green}{$\ast$} \\
\cline{1-5}
 45 & 0.55 & 19.5876 & 0.0490439 &\textcolor{red}{+}\\
 45 & 0.60 & 12.7990 & 0.0449062 &\textcolor{green}{$\ast$}\\
 45 & 0.65 & 11.2725 & 0.0458165 &\textcolor{green}{$\ast$}\\
 45 & 0.70 & 10.6459 & 0.0519231 &\textcolor{green}{$\ast$}\\
 45 & 0.75 & 10.3644 & 0.0668463 &\textcolor{green}{$\ast$}\\
\cline{1-5}
 50 & 0.55 & 15.1512 & 0.0447882 &\textcolor{blue}{$\times$}\\
 50 & 0.60 & 11.5498 & 0.0405102  & \textcolor{green}{$\ast$}\\
 50 & 0.65 & 10.5223 & 0.0408349  &\textcolor{green}{$\ast$}\\
 50 & 0.70 & 10.0892 & 0.0457436 &\textcolor{green}{$\ast$}\\
 50 & 0.75 & 9.90415 & 0.0582548 &\textcolor{green}{$\ast$}\\
\cline{1-5}
 55 & 0.55 & 12.7974 & 0.0405326 & \textcolor{green}{$\ast$} \\
 55 & 0.60 & 10.607  & 0.0361141 &\textcolor{green}{$\ast$}\\
 55 & 0.65 & 9.90434 & 0.0358532 &\textcolor{green}{$\ast$}\\
 55 & 0.70 & 9.61159 & 0.0395641 &\textcolor{green}{$\ast$}\\
 55 & 0.75 & 9.50027 & 0.0496634 &\textcolor{green}{$\ast$}\\
\cline{1-5}
 60 & 0.55 & 11.2825 & 0.0362769 &\textcolor{green}{$\ast$}\\ 
 60 & 0.60 & 9.86283 & 0.0317181 &\textcolor{blue}{$\times$}\\
 60 & 0.65 & 9.38387 & 0.0308716 &\textcolor{blue}{$\times$}\\
 60 & 0.70 & 9.19598 & 0.0333846  &\textcolor{blue}{$\times$}\\
 60 & 0.75 & 9.14208 & 0.0410719 & \textcolor{blue}{$\times$}
\\ \hline
\end{tabular}
\caption{Isosvector coupling constants $g_{\rho}$ and $\Lambda_{\omega}$ for
  $J=30$ and for different values of $L$. We also show whether the set of
  parameters gives a physical solution inside the $\chi$EFT band (marked as
  \textcolor{green}{$\ast$}), an unphysical one showing unstable regions
  (indicated as \textcolor{red}{+}) or outside the $\chi$EFT band (depicted by
  \textcolor{blue}{$\times$}). The analysis of the possible solutions is shown
  in Fig.~\ref{Slope30}.}
\label{label1}
\end{center}
\end{table}

  % ==============================================================================
%  Isovectors, just the physical ones...\\
\newpage 
\begin{table}[th]
\begin{center}
\begin{tabular}{|c|c|c|c|c|}
\hline
 \hspace{.2cm} L [MeV] \hspace{.2cm} & \hspace{.1cm} $m^{*}/m_N$
\hspace{.1cm} & \hspace{.2cm} $g_{\rho}$ \hspace{.2cm} & \hspace{.1cm}
$\Lambda_{\omega}$ \hspace{.1cm} & Solution \\
\cline{1-5}
 40 & 0.55 & 21.8033 & 0.0416554 &\textcolor{red}{+}\\
 40 & 0.60 & 14.5748 & 0.04026 &\textcolor{blue}{$\times$}\\
 40 & 0.65 & 12.7802 & 0.0429272 &\textcolor{blue}{$\times$}\\
 40 & 0.70 & 11.9943 & 0.0504747 &\textcolor{green}{$\ast$} \\
 40 & 0.75 & 11.6139 & 0.0670394 &\textcolor{green}{$\ast$} \\
\cline{1-5}
 45 & 0.55 & 17.2734 & 0.0386202 &\textcolor{red}{+}\\
 45 & 0.60 & 13.2057 & 0.0370172 &\textcolor{blue}{$\times$}\\
 45 & 0.65 & 11.9513 & 0.0391607&\textcolor{green}{$\ast$}\\
 45 & 0.70 & 11.3799 & 0.045714 &\textcolor{green}{$\ast$}\\
 45 & 0.75 & 11.107 & 0.0603224 &\textcolor{green}{$\ast$}\\
\cline{1-5}
 50 & 0.55 & 14.745 & 0.0355851 &\textcolor{red}{+}\\
 50 & 0.60 & 12.1617 & 0.0337743 & \textcolor{green}{$\ast$}\\
 50 & 0.65 & 11.2653 & 0.0353941 &\textcolor{green}{$\ast$}\\
 50 & 0.70 & 10.8512 & 0.0409534 &\textcolor{green}{$\ast$}\\
 50 & 0.75 & 10.6612 & 0.0536054 &\textcolor{green}{$\ast$}\\
\cline{1-5}
 55 & 0.55 & 13.0772 & 0.0325499 &\textcolor{blue}{$\times$}\\
 55 & 0.60 & 11.3317 & 0.0305315 &\textcolor{green}{$\ast$}\\
 55 & 0.65 & 10.6852 & 0.0316276 &\textcolor{green}{$\ast$}\\
 55 & 0.70 & 10.3899 & 0.0361928 &\textcolor{green}{$\ast$}\\
 55 & 0.75 & 10.2651 & 0.0468884 &\textcolor{green}{$\ast$}\\
\cline{1-5}
 60 & 0.55 & 11.8716 & 0.0295148 &\textcolor{green}{$\ast$}\\
 60 & 0.60 & 10.6513 & 0.0272887 &\textcolor{green}{$\ast$}\\
 60 & 0.65 & 10.1865 & 0.0278611 &\textcolor{green}{$\ast$}\\
 60 & 0.70 & 9.9829  & 0.0314321 &\textcolor{green}{$\ast$}\\
 60 & 0.75 & 9.9101  & 0.0401714 &\textcolor{green}{$\ast$}
\\ \hline
% \hline
\end{tabular}
\caption{The same as Table~\ref{label1}, but for $J=32$. The analysis of the
  possible solutions is shown in Fig.~\ref{Slope32}.}
\label{label11}
\end{center}
\end{table}
% &&&&&&&&&&&&&&&&&&&&&&&&&&&&&&&&&&&&&&&&&&&&&&&&&&&&&&&&&&&&&&&&&&&&&&&&&&&&&&&&&&&&&&&&&&&&&&&&&&&&&&&&&
% &&&&&&&&&&&&&&&&&&&&&&&&&&&&&&&&&&&&&&&&&&&&&&&&&&&&&&&&&&&&&&&&&&&&&&&&&&&&&&&&&&&&&&&&&&&&&&&&&&&&&&&&&

% \newpage

\begin{acknowledgments}
  L.T. acknowledges support from the Ram\'on y Cajal research programme,
  FPA2013-43425-P and FPA2016-81114-P Grants from Ministerio de Econom\'{\i}a
  y Competitividad (MINECO), Heisenberg Programme of the Deutsche
  Forschungsgemeinschaft under the Project Nr. 383452331 and PHAROS COST
  Action CA16214. J.S. and A.Z. acknowledge support from the Hessian LOEWE initiative
  through the Helmholtz International Center for FAIR (HIC for FAIR).
\end{acknowledgments}

\bibliography{neue_bib_mod}
\bibliographystyle{apsrev4-1}
\end{document}